\documentclass[fleqn,10pt,usenames,dvipsnames]{wlscirep}
\usepackage[T1]{fontenc}
\usepackage[utf8]{inputenc}
\usepackage{authblk}
\usepackage{float}
\usepackage{graphicx}
\usepackage{xcolor}
\usepackage{colortbl}
\usepackage{wrapfig}
\usepackage{capt-of}
\usepackage{hyperref}
\usepackage{fullpage}

\title{Transition Matrices: A Tool to Assess Student Learning and Improve Instruction}

\author[1,*]{Gary A.~Morris}
\author[1,+]{Paul J.~Walter}
\author[2]{Spencer Skees}
\author[3]{Samantha Schwartz}
\affil[1]{St.~Edward's University, Mathematics Department, Austin, TX, USA}
\affil[2]{MSEED Program, Valparaiso University, Valparaiso, IN, USA}
\affil[3]{Plymouth High School, Plymouth, IN, USA}
\affil[*]{garymorris@stedwards.edu}
\affil[+]{pauljw@stedwards.edu}

\begin{document}

\begin{abstract}
This paper introduces a new spreadsheet tool for adoption by high school or college level physics teachers who use common assessments in a pre-instruction/post-instruction mode to diagnose student learning and teaching effectiveness.  The spreadsheet creates a simple matrix that identifies the percentage of students who select each possible pre-/post-test answer combination on each question of the diagnostic exam.  Leveraging analysis of the quality of the incorrect answer choices, one can order the answer choices from worst to best (i.e., correct), resulting in ``transition matrices'' that can provide deeper insight into student learning and the success or failure of the pedagogical approach than traditional analyses that employ dichotomous scoring. 
\end{abstract}

\flushbottom
\maketitle
%
%
\thispagestyle{empty}

\section*{The Force Concept Inventory}

Among a number of important assessment instruments emerging from the PER community\cite{bardar2007development,ding2006evaluating,hestenes1992mechanics,maloney2001surveying,thornton1998assessing}---a more complete list is available at PhysPort \cite{physport}---the Force Concept Inventory (FCI)\cite{hestenes1992force} has become the most widely used test of student knowledge and physics instruction quality in Newtonian mechanics.  We focus our analysis in this paper on college-level student data from the FCI in the sections of the Rice University calculus-based, pre-medical version of the introductory physics course taught by the lead author (2000 – 2003).  The version of the FCI used in this study appears in Mazur’s Peer Instruction \cite{mazur1997user} and includes 30 multiple-choice questions with five answer choices for each question.

The FCI was designed to incorporate common misconceptions, so the incorrect answer choices attract students lacking a full understanding of the concepts. Dedic et al.\cite{dedic2010all}~provides evidence that response patterns on the FCI form seven reasoning-groups.  Transitions between these groups are correlated with the level of understanding of the students, with a clear hierarchy among the groups in evidence.  They posit that the incorrect answers on the FCI are not equivalent.  This finding was confirmed by Morris et al.\cite{morris2012item}~using the item response curves analysis (IRC)\cite{morris2006testing} to identify the functionality of items on the FCI and to correlate particular answer choices with particular ranges of student understanding.  They found that for most questions on the FCI, a sequence of incorrect answer choices could be identified, ordered from the lowest to the highest levels of understanding.  Taken together, these two studies suggest that some of the attractive incorrect answer choices on the FCI represent intermediate steps on the progression from a student ignorant of the physics content to one who has mastered it.

\begin{figure}
\centering
\begin{minipage}{.5\textwidth}
  \fbox{
  \begin{minipage}{0.62\linewidth}
    \noindent 18.~The figure below shows a boy swinging on a rope, starting at a point higher than A.   Consider the following distinct forces:
    \vspace{-.2cm}
    \begin{enumerate}
      \item A downward force of gravity. \vspace{-.2cm}
      \item A force exerted by the rope pointing from A to O. \vspace{-.2cm}
      \item A force in the direction of the boy’s motion. \vspace{-.2cm}
      \item A force pointing from O to A.\vspace{-.2cm}
    \end{enumerate}
    Which of the above forces is (are) acting on the boy when his is at position A?
    \vspace{-.2cm}
    \begin{description}
      \item[(A)] 1 only.\vspace{-.2cm}
      \item[(B)] 1 and 2.\vspace{-.2cm}
      \item[(C)] 1 and 3.\vspace{-.2cm}
      \item[(D)] 1, 2, and 3.\vspace{-.2cm}
      \item[(E)] 1, 3, and 4.
    \end{description}
  \end{minipage} \hfill
  \begin{minipage}{0.3\linewidth}
    \includegraphics{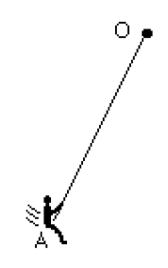}
  \end{minipage}
  } 
  \captionof{figure}{Question 18 on the Force Concept Inventory (FCI).\cite{hestenes1992force} Reproduced with permission of Dr.~David Hestenes.}
  \label{fig:Q18-FCI}
\end{minipage}%
\hspace{.2cm}
\begin{minipage}{.45\textwidth}
  \centering
  \includegraphics[width=\linewidth]{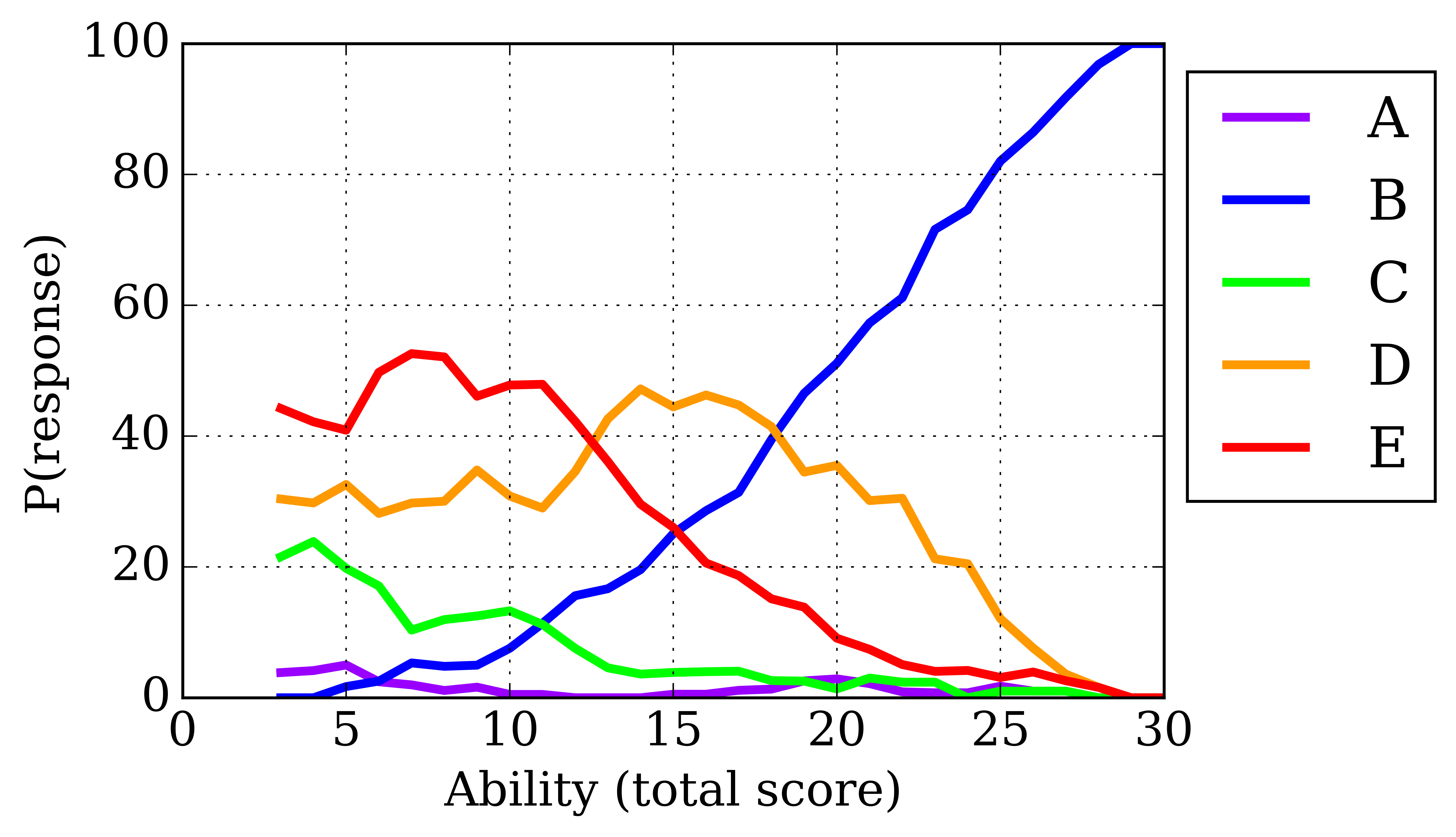}
  \captionof{figure}{Item response curves (IRCs)\cite{morris2006testing} for Question 18 of the Force Concept Inventory (FCI)\cite{hestenes1992force} using data from students at Rice University taught by the lead author (2000 -- 2003).  The different curves correspond to the five different answer choices for this item.  The percent of students selecting each answer choice is plotted as a function of total score on the FCI, a proxy for student ability (a good approximation for the FCI, as shown by Wang and Bao\cite{wang2010analyzing}).   Data are smoothed using a simple 3-bin moving average (one-sided 2-bin average for the endpoints).  Data with total scores $< 3$ are not shown given the small sample size.}
  \label{fig:Q18-IRCs}
\end{minipage}
\end{figure}

Figure \ref{fig:Q18-FCI} is a reproduction of Question 18 on the FCI.  Figure \ref{fig:Q18-IRCs} provides an example of IRCs computed for Question 18 on the FCI using data from $\sim 500$ Rice student tests.  Each IRC relates the percentage of students at a given level of understanding (total score) that selects each answer choice.  As shown in Wang and Bao\cite{wang2010analyzing}, total score on the FCI is strongly correlated with understanding of the material.

\section*{Construction of the Transition Matrix}

We leverage the well-designed multiple choice items on the FCI to develop the idea of a transition matrix---a simple, analytic tool that enables physics instructors to close the loop on assessment.  Transition matrices can be thought of as an adaptation of a two-observation/one-link Markov chain, where the sum of the all of the elements in the matrix is 100\%. 

Table \ref{tab:trans-matrix} presents an example transition matrix corresponding to the pre- and post-instruction data from Rice University students responding to Question 18 on the FCI.  Each element in the transition matrix represents the percentage of students who select a given answer choice to a specific item on the pre-test (the rows) and a given answer choice to that same item on the post-test (the columns).  The columns and rows of the transition matrix are ordered with the ``worst'' incorrect answer choice in the top row and leftmost column, and the correct (best) answer choice in the bottom row and rightmost column.  The intermediate rows and columns are ordered here as suggested by the IRCs in the 6000+ test database of Morris et al.\cite{morris2012item}  The IRCs are constructed using data from University students.  While differences may exist if IRCs are constructed using data from other populations (e.g., high school students), the statistics used to construct the IRCs should become stationary with sufficiently large populations under the assumption that the relationship between particular answer choices and ``Newtonian thinking'' is independent of background, level of study, demographics, and other population-specific measures (proof of this assumption is beyond the bounds of this paper and will be addressed in a future analysis).  Further, by ranking the answer choices using the IRCs, we are not guaranteed that better answer choices represent being closer to Newtonian thinking.   
\begin{wraptable}[14]{l}{0.52\textwidth}
    \begin{tabular}{|c|c|c|c|c|c|}
      \hline
      \multicolumn{6}{|c|}{\textbf{Question 18 (Answer = B)}}\\ \hline
      & \multicolumn{5}{c|}{\textbf{Post}}\\ \hline    
      \textbf{Pre}   & \textbf{A} & \textbf{C} & \textbf{E} & \textbf{D} & \textbf{B} \\ \hline
      \textbf{A}   & \cellcolor[HTML]{FFB9B9} 0.1\% & \cellcolor{YellowGreen}0.1\% & \cellcolor{YellowGreen}0.2\% & \cellcolor{YellowGreen}0.4\% & \cellcolor{green} 0.4\% \\ \hline
      \textbf{C}   & \cellcolor{BrickRed}0.2\% & \cellcolor[HTML]{FFB9B9}1.0\% & \cellcolor{YellowGreen}0.8\% & \cellcolor{YellowGreen}2.8\% & \cellcolor{green}3.6\% \\ \hline
      \textbf{E}   & \cellcolor{BrickRed}1.0\% & \cellcolor{BrickRed}1.2\% & \cellcolor[HTML]{FFB9B9}7.4\% & \cellcolor{YellowGreen}18.4\% & \cellcolor{green}14.5\% \\ \hline
      \textbf{D}   & \cellcolor{BrickRed}0.4\% & \cellcolor{BrickRed}0.6\% & \cellcolor{BrickRed}1.6\% & \cellcolor[HTML]{FFB9B9}9.9\% & \cellcolor{green}18.0\% \\ \hline
      \textbf{B}   & \cellcolor{red} 0.1\% & \cellcolor{red} 0.2\% & \cellcolor{red} 0.6\% & \cellcolor{red} 1.8\% & \cellcolor{Green} 13.5\% \\ \hline
    \end{tabular}
\captionof{table}{The transition matrix for Question 18 on the Force Concept Inventory (FCI)7 using data from students at Rice University taught by the lead author from 2000 – 2003. } 
\label{tab:trans-matrix}
\end{wraptable}
In Dedic et al.\cite{dedic2010all}, students were more likely to transition to the best reasoning-group (Class 1—representing Newtonian thinking) from the third best reasoning group (Class 3) rather than the second best reasoning group (Class 2). In the transition matrix approach, the transition rates for each answer choice of a question are embedded in the transition matrix; higher rates of transitioning to the correct answer may come from a worse wrong answer than a better wrong answer.   Rather the ranking produced from the IRCs simply relate the likelihood of selecting particular answer choices to the total score on the FCI, the latter of which corresponds to an evaluation of the level of ``Newtonian thinking'' of the student.
\enlargethispage*{1em}
\pagebreak

\section*{Interpreting the Transition Matrix}
In the transition matrix, a perfect learning outcome would result in 0s in all entries except those of the last column, with the sum of the entries in the last column totaling 100\%.  At the other end of the spectrum, in a population in which no learning took place (and students did not guess), all elements of the matrix would be 0 except those on the diagonal (color coded in pink in Table \ref{tab:trans-matrix}), whose sum would be 100\%.  For an instructor, the latter result would be very troubling, as it would suggest that instruction made no difference in the class' level of understanding of the concept tested in that item.  The color-coding of the transition matrix (Table \ref{tab:trans-matrix}) enables instructors to quickly gauge changes in their students' understanding.  Elements above the diagonal (color coded in green) represent the students who moved to better answers after instruction; elements below the diagonal (color coded in deep red) represent the students who moved to worse incorrect answers after instruction.  This transition matrix thus provides a great deal more information to the instructor than dichotomous scoring, gain (pre-/post-test score difference), or the Hake normalized gain.\cite{hake1998interactive}

\section*{Example Analysis of the Transition Matrices from a Question on the FCI}
 
Table \ref{tab:matrix-results} is a derivative of the transition matrix in Table \ref{tab:trans-matrix} that provides a summary of the results on Question 18.  It summarizes the transitions from an initial incorrect answer to the correct answer, a better incorrect answer, the same incorrect answer, or a worse incorrect answer (from top to bottom).  The instructor might not be heartened by the fact that only 50\% of students arrived at the correct answer post instruction (the sum of the entries in the right most column of Table \ref{tab:trans-matrix}).  The summary table for the transition matrix, however, shows that another 22.7\% of students moved to a better incorrect answer choice after instruction.  One might argue that these students moved part of the way but did not quite arrive at a complete understanding.  Another positive observation from the matrix is that only 7.7\% of students who got the item incorrect initially moved to a worse answer choice after instruction, while 72.7\% of those who got it incorrect initially got it correct or moved to a better incorrect answer after instruction. Such detailed analysis of student performance provides a richer assessment of student learning outcomes and teaching effectiveness.\\

\begin{minipage}{.8\textwidth}
 
  \begin{tabular}{|c|c|c|c|c|}
      \hline 
      \textbf{Question 18}   & \textbf{summary} & \textbf{Transition} & \textbf{raw} & \textbf{normalized} \\ \hline
      \textbf{pre correct}   & 16.2\% & \cellcolor{Green} correct $\rightarrow$ correct & 13.5\% & 83.3\% \\ \hline
      \textbf{pre incorrect}   & 82.6\% & \cellcolor{green} incorrect $\rightarrow$ correct & 36.5\% & 44.2\%  \\ \hline
       &  & \cellcolor{YellowGreen} incorrect $\rightarrow$ better incorrect & 22.7\% & 27.5\%  \\ \hline
      \textbf{post correct}   & 50.0\% & \cellcolor[HTML]{FFB9B9} kept incorrect & 18.4\% & 22.3\%  \\ \hline
      \textbf{post incorrect}   & 50.0\% & \cellcolor{BrickRed}incorrect $\rightarrow$ worse incorrect & 5.0\% & 6.1\%  \\ \hline
      & & \cellcolor{red} correct $\rightarrow$ incorrect & 2.7\% & 16.7\% \\ \hline
    \end{tabular}
\captionof{table}{Summary table for the transition matrix of Table \ref{tab:trans-matrix}.  At left are the fractions of students who got Question 18 on the FCI correct or incorrect before (pre) and after (post) instruction.}
\label{tab:matrix-results}
\end{minipage}
\vspace{1em}

Different pedagogical approaches might be required to move students along the progression from one incorrect answer choice to the next better one (i.e., moving from the worst to the second worst answer choice requires a different approach than moving from the best incorrect answer choice to the correct answer choice).  Indeed Table \ref{tab:trans-matrix} shows that the instructor was more successful in moving his students from answer choice E to answer choice D (the second best to the best incorrect answer choice), at a rate of 18.4\%, than in moving them from answer choice E to the correct answer choice B, at a rate of 14.5\% of the total population.  Said another way, 43\% of students who started with answer choice E moved to answer choice D while only 34\% moved to the correct answer choice. By considering not just the fraction of students who get this item correct but also the hierarchy of answer choices, we see an improvement in student understanding after instruction that was not apparent from the traditional dichotomous scoring approach. The most persistent incorrect answer, choice D, deals with the misconception that a force is needed to keep the boy in the problem moving. While this ``force'' is incorrect, this answer choice represents an improvement from the pre-test selection of answer choice E, which states that a force is needed to keep the boy moving and confuses centripetal and centrifugal force. The instructor still might ask what he/she could do differently the following year to help more students move from answer choice E to the correct answer choice.

\section*{Concept and Misconception Analysis}

In addition to the transition matrices described above, the spreadsheet tool also provides instructors feedback on the concepts their students have mastered and misconceptions that are most common among their students as based on the taxonomies found in Hestenes et al.\cite{hestenes1992force}~(using the 1995 revised Tables I and II).  Such a report, used prior to instruction, could help instructors identify those topics on which they should spend the most time during class as well as those already mastered by their students, on which they need not spend much time (if any).  Used after instruction, instructors could quickly identify areas that either the text or their lesson plans led to adoption more strongly of particular misconceptions, which then allows the instructor to revise his/her lesson plans to mitigate or correct any introduced confusion.

\section*{Available Resource for Instructors}
A spreadsheet that computes the transition matrices based on item-level data entered by the instructor is available on the American Modeling Teachers Association webpage at  \href{https://modelinginstruction.org/effective/evaluation-instruments/fci-transition-matrix-tool/}{https://modelinginstruction.org/effective/evaluation-instruments/fci-transition-matrix-tool/}. 

\section*{Adopting Transition Matrices Approach for Another Assessment}

We also note that the basic tool could be adapted for use with other diagnostic instruments and within disciplines other than physics.  For each question, construct the item response curves (IRCs), rank the answer choices, and then construct the transition matrix.  Discernable from the IRCs, the transition matrices approach requires that the wrong answers function as distractors---rather than being random answer choices.  If the new assessment has 5 answer choices per question, then one only needs to update the ordering of the answer choices on the spreadsheet; a different number of answer choices will require programming changes to the spreadsheet macro. 

\section*{Summary}

We have introduced an openly available tool that practitioners can easily implement with a widely used diagnostic (i.e., the FCI) to improve instruction.  The tool enables instructors to receive feedback not only at the beginning of the semester based on which concepts and misconceptions require the most attention, but also from one iteration of the course to the next based on the transitions in understanding of Newtonian mechanics their students are making as the result of instruction.  Secondarily, this paper introduces an approach that can be adapted to other diagnostics in physics and other fields.  

\section*{Acknowledgements}

Student researcher funding for this project was provided by the National Science Foundation DUE STEP Grant Award \#1068346.  Special thanks to Eric Mazur (Harvard University) and Taha Mzoughi (Kennesaw State University) for data from Harvard and Mississippi State respectively that led to the ranking of incorrect answer choices for the FCI used in this paper.  Thanks also to Lee Branum-Martin (Georgia State University), Nathan Harshman (American University), Steven Baker (Rice University), and Kelly Miller (Harvard University) for useful discussions in the development of this tool.  Thanks to Brenna Thompson (St.~Edward’s University) for her work in checking the tool.  Finally, thanks to Prof.~Jan Westrick in the Dept.~of Education at Valparaiso University who provided valuable feedback on an early version of this paper and whose suggestions greatly improved this paper.

\bibliography{references}

\end{document}